\documentstyle [11pt] {article}
 \newcommand {\be}{\begin{equation}}
 \newcommand {\ee}{\end{equation}}
 \newcommand{\ls}{\;\raisebox{-.8ex}{$\buildrel{\textstyle<}\over\sim$}\;}
 \newcommand{\anrev}{{\it ARA\&A, }}
 \newcommand{\apj}{{\it ApJ, }}
 \newcommand{\apjs}{{\it ApJS, }}
 \newcommand{\aj}{{\it AJ, }}
 \newcommand{\mnr}{{\it MNRAS, }}
 \newcommand{\nat}{{\it Nat, }}

 \newcommand{\ana}{{\it A\&A, }}

\begin{document}
 \title{{\bf VARIABLE SMOOTHING LENGTHS AND ENERGY CONSERVATION  IN
 SMOOTHED PARTICLE HYDRODYNAMICS.}}

 \author{Richard P. Nelson \& John C. B. Papaloizou,\\ \\
 Astronomy Unit,\\ \\
 School of Mathematical Sciences,\\ \\
 Queen Mary \& Westfield College,\\ \\
 Mile End Road,\\ \\
 London E1 4NS.}

 \maketitle
 \newpage

\noindent{\bf Abstract.}
 We present a new formulation of the equations of motion used in smoothed
 particle hydrodynamics (SPH). The spatial resolution in SPH is determined
 by the smoothing length, $h$, and it has become common practice for
 each particle to be given its own adaptive smoothing length, $h_i$. This has
 the advantage that the dynamic range that may be spatially resolved is greatly
 increased,
 but also has the drawback that additional terms which account for the
 variability of the smoothing lengths should be included in the particle
 equations of motion in order to satisfy conservation requirements.
 We refer to these additional terms as $\nabla h$ terms. The difference
 between our approach and that of previous implementations is that
 these $\nabla h$ terms have now been included in the equations of motion,
 whereas they had previously been neglected. This is achieved by
 defining a functional form for the $h_i$s, that depends on
 inter--particle distances only, and then deriving the equations of motion
 using a Hamiltonian formalism.

 A number of test calculations, designed to compare the effects of
 including and ignoring these $\nabla h$ terms, are
 presented. We find that the inclusion of the $\nabla h$ terms has no
 detrimental effects on the ability of SPH to model known problems,
 such as one-dimensional shock--tubes or the adiabatic collapse of
  cold gas spheres, with reasonable accuracy, and may in some
 cases lead to improvements in
 their qualitative outcome. For problems on which SPH has been shown
 to perform rather badly, because of poor energy conservation, we find
 that including the $\nabla h$ terms results in a dramatic improvement.
 In particular, non--conservation of energy during a head on collision
 between identical polytropes can occur at the level of $\Delta E \approx
 10 \%$ when the $\nabla h$ terms are neglected. When the $\nabla h$ terms
 are included, however, then this error reduces to $\Delta E \approx 0.8. \%$

\section{Introduction.}

 Smoothed particle hydrodynamics was introduced in order to study problems
 in astrophysics involving the motion of compressible fluid masses of
 arbitrary geometry in
 three dimensions (Lucy 1977; Gingold \& Monaghan 1977). For a recent review
 see Monaghan (1992), and references therein. Particles are used to
 represent a sub-set of the fluid elements that arise
 in the Lagrangian description of a fluid, and because spatial derivatives
 are calculated analytically from interpolation formulae, rather than on
 a grid, SPH is by its very
 nature adaptive and thus well suited to problems in which large density
 contrasts can occur, such as the fragmentation of self-gravitating gas clouds.

 The spatial resolution of SPH is determined by the smoothing length, $h$.
 In  early formulations $h$ was either taken to be constant, or else
 was a globally time-dependent function of the mean number density
 of particles in the system. More recently, however, it has become
 common for each particle to have its own time dependent smoothing length
 which adapts according to the local number density of particles.
 Full advantage is then taken of the Lagrangian nature of SPH,
 and the dynamic range in spatial resolution of the method is dramatically
 increased.
 The introduction of time varying smoothing lengths can, however, lead to
 serious problems with energy conservation in certain situations, as was
 highlighted in a recent paper by Hernquist (1993).
 The problem arises from the fact that the use of variable smoothing
 lengths means that additional terms should appear in the particle equations
 of motion. Up until recently, these additional terms have been ignored, and
 as a consequence the particle equations of motion are no longer conservative.

 In a previous paper (Nelson \& Papaloizou 1993), using a Hamiltonian
 formalism, we have derived a set of conservative
 equations for a barotropic fluid which include terms
 accounting for the variability of the smoothing lengths.
 In this
 paper these equations are generalised for the case of an
 adiabatic fluid in which
 the entropy on each particle is conserved. The treatment is then
 extended to include the effects of viscous dissipation.

 By means of simple numerical tests, a comparative study
 between the newly derived version of SPH and a more standard formulation of
 the method, including spatially and temporally varying smoothing lengths, is
 presented. An initial set of calculations is presented, in order to compare
 the qualitative outcome of different versions of our code when it is used
 to model known problems, such as one-dimensional Riemann shock-tubes, and
 the adiabatic collapse of an initially isothermal gas sphere. The inclusion
 of additional terms in the particle equations of motion are not found to have
 any detrimental effects on the qualitative outcome of the calculations, and in
 some cases can lead to improvements as a result of
 more accurate energy conservation.
 A second set of test calculations is presented that is specifically
 designed to test the conservation properties of the SPH algorithm. It is
 found that errors in the conservation of energy or entropy can occur at the
 level of $\approx 10 \%$, when the additional $\nabla h$ terms are neglected,
 during a head-on collision between identical polytropes. For the same
 calculation, it is found that
 these errors are reduced to $\approx 0.8 \%$ when the
 $\nabla h$ terms are included.

 The basic hydrodynamical and thermodynamical equations of the problem
 are presented in Section (2). The forms of these equations used in SPH
 methods
 are derived and discussed in Section (3). In Section (4), calculations
 designed to compare the qualitative performance of different versions
 of the SPH method are described. In Section (5) we describe and discuss
 the results from calculations designed to test the conservation properties
 of the different formulations. Calculations designed
 to test the effect of the $\nabla h$ terms on the numerical diffusivity of
 SPH are presented in Section (6). Finally, conclusions are drawn in
 Section (7),
 and we present a general discussion on issues arising from
 our calculations, and on the interpretation of the $\nabla h$ terms.

\section{Equations of motion.}
 The  equations of motion for a compressible fluid can be
 written

\be  \frac{d\rho}{dt} + \rho\nabla \cdot {\bf v} = 0 \label{cont} \ee

\be \frac{d{\bf v}}{dt} = - \frac{1}{\rho} \nabla P - \nabla\Phi +
 {\bf{S}}_{visc} \label{dvdt}
  \ee

 \be \rho \frac{d {\cal U}}{dt} + P \nabla \cdot {\bf v} = \rho \cal H
\label{dudt}  \ee where
$${d\over dt}\equiv {\partial \over\partial t}+{\bf v}\cdot\nabla$$
denotes the convective derivative,
 $\rho$ is the density, ${\bf
v}$ the velocity, P the pressure, ${\cal U}$ the internal energy per unit mass,
and $\cal H$ a term comprising all non-adiabatic
heating and cooling rates per unit mass. This set of equations is
supplemented by an
 equation of state, which  can be written in the form

\be P \equiv P\left({\cal U}, \rho \right)\label{state1} \ee
 in which the pressure  is written as a function of the internal energy
per unit mass and the density, which may be adopted as the two thermodynamic
variables defining the state of the fluid.

\noindent Alternatively, we may introduce the
entropy per unit mass, $S$,
which satisfies the equation
\be  T \frac{d S}{dt} =  \cal H \label{dSdt} \ee
 where $T$ is the temperature. The quantities
$T, P,$ and ${\cal U}$ are then expressed as functions of
the two thermodynamic variables $\rho$ and $S.$
We note the thermodynamic relation
$$T=\left.{\partial {\cal U}\over \partial S}\right|_{\rho}.$$

\noindent For the special case considered here
of an ideal gas with constant specific
heat capacity at constant volume $c_v$, and constant
specific heat ratio $\gamma ,$  we have
\be P = (\gamma -1)\rho {\cal U}, \label{state2} \ee
\be P = c_v(\gamma-1)\rho T, \label{state3} \ee
and
\be S = c_v \log\left( T \rho^{1-\gamma}\right). \label{state4} \ee
In this case, it is also sometimes convenient to work
in terms of the quantity
$$K={P\over \rho^{\gamma}}, \ \ {\rm where} $$
$$K=c_v(\gamma - 1)\exp\left( {S\over c_v}\right)$$
and is sometimes referred to as the entropic function.

 \section{Numerical method.}
 The above set of equations is solved numerically using SPH
(Lucy 1977; Gingold \& Monaghan 1977). As described above,
 SPH uses particles to represent a sub-set of the fluid elements that arise in
 the Lagrangian description of a fluid. This set of particles is chosen in
 such a way as to ensure that the particles' mass distribution maps the fluid
 density, $\rho$. By calculating the trajectories of the particles according to
 the hydrodynamic equations, it is then possible to calculate the properties of
 the fluid at later times.

 \subsection{Hydrodynamical equations.}
 \subsubsection{Density.}
 In order to reduce the statistical fluctuations arising from representing a
 fluid continuum by a finite set of particles, it is necessary to introduce
 a smoothing procedure. A smoothed quantity may be  defined by the expression
  \be \langle f({\bf r})\rangle = \int f({\bf r'}) W(|{\bf r}
  - {\bf r'}|,h) d^{3}{\bf r'} \label{sf1} \ee
 where $f({\bf r})$ is an arbitrary function, $W$ is a smoothing kernel, and
 $h$ is the smoothing length.
 For a function $f({\bf r})$ defined only at a discrete set of points,  a
 Monte Carlo representation of equation (\ref{sf1}) may be written

 \be \langle f({\bf r}_{i}) \rangle = \sum^{N}_{j=1}\frac{m_{j}}{\rho_{j}
 }
  f({\bf r}_{j}) W(|{\bf r}_{i} - {\bf r}_{j} |, h)
  \label{sf2} \ee
 where $m_{j}$ is the mass of particle $j$, and $\rho_{j}$ is the density at
the location of
 particle $j$.
 If the function $f({\bf r})$ is chosen to be the density, $\rho({\bf r})$,
 then we obtain

 \be \rho({\bf r}_{i}) = \sum_{j=1}^{N} m_{j} W(|{\bf r}_{i} - {\bf r}_{j}|, h)
  \label{den1} \ee
 where the angled brackets have been dropped for the sake of brevity, with the
 understanding that smoothed functions are used throughout in SPH.

\noindent The spatial resolution of a calculation is determined by the
 smoothing length, $h$.
 Most implementations of the SPH algorithm employ temporally and spatially
 variable smoothing lengths in order to increase the dynamic range that can be
 spatially
 resolved, and hence take full advantage of the Lagrangian nature of the
method.
 When variable smoothing lengths are employed each particle, $i,$ will have an
associated smoothing length $h_{i}.$ In the standard formulation of SPH, it is
 then necessary, in order to conserve momentum,
 to symmetrise  expressions
 such as $$ W(|{\bf r}_{i} - {\bf r}_{j}|, h) \equiv W_{ij}$$
 with respect to $i$ and $j$.
 In Nelson \& Papaloizou (1993) we set, following  (Evrard 1988),
 \be h \rightarrow h_{ij}= \frac{h_{i}+h_{j}}{2}. \label{hij} \ee
 Alternatively one can  use the method suggested
 by Hernquist \& Katz (1989) which symmetrises using the replacement

 \be W_{ij} \rightarrow \frac{1}{2}[ W(|{\bf r}_{i} - {\bf r}_{j} |, h_{i}) +
 W(|{\bf r}_{i} - {\bf r}_{j} |, h_{j})]. \label{Wij} \ee
 For comparison purposes,
 we have adopted the method
 of Hernquist \& Katz (1989) in this paper.
 The smoothed density (\ref{den1}) then becomes

 \be \rho({\bf r}_{i}) = \sum_{j=1}^{N} m_{j} \frac{1}{2}[ W(|{\bf r}_{i}
 - {\bf r}_{j} |, h_{i}) + W(|{\bf r}_{i} - {\bf r}_{j} |, h_{j})]
 \label{den2}  \ee
 \subsubsection{Equations of motion.}
We now have to form equations of motion for each particle. From equation
(\ref{dvdt}) we obtain, denoting quantities associated with particle $i$
by a subscript $i,$

\be m_i\frac{d{\bf v}_i}{dt} = {\bf F}_ {P,i} + {\bf F}_ { G, i}+
 {\bf{S}}_{visc, i} . \label{dvdti}
  \ee
Here we have denoted the forces  on particle $i$ due to pressure,
gravity and viscosity by ${\bf F}_ {P, i} , {\bf F}_ {G, i} ,$ and
 ${\bf{S}}_{visc, i}$ respectively.

\subsubsection{A conservative system.}
In the early formulations of SPH  a
global smoothing length was used that could be either constant
(Lucy 1977), or a function
of time by relating it to the mean density of the
system (Gingold \& Monaghan 1977).
In an attempt to
increase the dynamic range that can be modelled, later formulations
(Evrard 1988; Hernquist \& Katz 1989)
have replaced the global smoothing length by one with a
spatial and temporal dependence, together with an appropriate symmetrisation
with respect to particle pairs (see above).
 This has the  consequence that, for barotropic or isentropic systems, the
particle equations of motion are no longer conservative as they should be
because the forces have not been derived from an appropriate potential
function.
It has been
 suggested in the literature (Gingold \& Monaghan 1982; Evrard 1988)
 that effects arising
from the variation
 of the $h_{i}$s in time and space should be small, although there is no firm
evidence to show that these terms may be
 neglected.
 This concern is also true for a system in which a time varying, global
 smoothing length is employed, as shown recently for
 example by Hernquist (1993),
 but it is expected that, in general, the effects of
 non-conservation will be less severe. This is because the time change in the
 smoothing length, which drives the non-conservation of energy, is attenuated
 by the fact that it is dependent on the global properties of the system
 rather than on purely local ones.

\noindent In his recent paper, Hernquist (1993) has shown
 that, under certain circumstances,
 errors at the $ \approx 10 \% $ level can occur in the conservation of
 either energy or entropy, depending on whether the entropy or energy
 equations are integrated, when modelling adiabatic flows. These errors were
 shown to arise as a direct consequence of the use of variable smoothing
 lengths. In view of this, it is desirable that a conservative formulation for
 the various terms in the particle equations of motion be derived that can
incorporate
  variable smoothing lengths. This has already been given for the pressure
force in a barotropic fluid in Nelson \& Papaloizou (1993). Here
we generalise the approach to the case of a general adiabatic flow
in which there may not be a barotropic equation of state, but in which the
entropy of each particle is conserved. We then extend the treatment to include
viscosity.

 \subsubsection{The pressure force.}
 An inviscid fluid with a barotropic equation of state is a Hamiltonian
 system, so it is desirable, when modeling such a system, that
 the particle equations of motion in SPH
 form a Hamiltonian system also. The thermal contribution to the total energy
 can be written (see Nelson \& Papaloizou, 1993) as

 $$ U = \int_{V} F(\rho) \rho d^{3}{\bf r}, \ {\rm where} \ {dF\over d\rho} =
 {P\over\rho^2}. $$
 This has the Monte Carlo representation

 \be U=\sum^{N}_{k=1}m_kF(\rho_k). \label{poten} \ee
 It is then natural to use $U$ as a potential function
 for deriving the pressure
 force on particle $i$ through
 $$ {\bf F}_{P, i} = - {\partial U\over \partial {\bf r}_i}. $$

\noindent The same formulation goes through for an adiabatic system for which
the entropy
of each particle is conserved. One simply replaces $F(\rho)$ in the above
by the internal energy per unit mass ${\cal U}(S,\rho)$ written as
a function of the entropy and the density. The Monte Carlo representation
for the pressure potential function is then
 \be U=\sum^{N}_{k=1}m_k{\cal U}_k(S_k,\rho_k), \label{poten1} \ee
where $S_k$ denotes the entropy per unit mass for particle $k.$
We remark that the property of ${\cal U}$ that is needed is
$$\left({\partial {\cal U}\over \partial \rho} \right)_{S}=
 {P\over\rho^2}. $$

\noindent For an adiabatic flow $S_k$ is constant, so that in both the
adiabatic
and barotropic cases
 the pressure force on particle $i$ is found, after differentiating either
(\ref{poten}) or (\ref{poten1}) and  using (\ref{den2}), to be
 \be {\bf F}_{P,i}=  - \sum_{k=1}^{N}\sum_{j=1}^{N}
  m_{k}m_{j}\left(\frac{P_{k}}{\rho^{2}_{k}} \right)
  \frac{\partial}{d{\bf r}_{i}}\frac{1}{2} \left[
 W(|{\bf r}_{k}-{\bf r}_{j}|,h_{k})+ W(|{\bf r}_{k}-{\bf r}_{j}|,h_{j}) \right]
 \label{gradp2} \ee
 which may be written in the form

 \begin{eqnarray}
  {\bf F}_{P,i} & = & - \sum_{j=1}^{N} m_{i} m_{j}
 \left(\frac{P_{i}}{\rho_{i}^{2}}+ \frac{P_{j}}{\rho_{j}^{2}} \right)
\frac{1}{2
}
 \left[
 \left.\frac{\partial W({\bf r}_{ij},h_i)}{\partial {\bf r}_i}
 \right|_{h_i const} +
 \left.\frac{\partial W({\bf r}_{ij},h_j)}{\partial {\bf r}_i}
  \right|_{h_j const}
  \right]
 \nonumber  \\
 & &  - \sum_{k=1}^{N} \sum_{j=1}^{N} m_k m_j
 \left(\frac{P_{k}}{\rho_{k}^{2}} \right) \frac{1}{2} \left[
 \frac{\partial W({\bf r}_{kj},h_k)}{\partial h_k}  \frac{\partial h_k}
 {\partial
 {\bf r}_{i}} + \frac{\partial W({\bf r}_{kj},h_j)}{\partial h_j}
 \frac{\partial h_j}{\partial{\bf r}_{i}} \right] \label{gradp3}
 \end{eqnarray}
 where we adopt the notation ${\bf r}_{kj} \equiv {\bf r}_k - {\bf r}_j.$
The first part of the above pressure force, which does not involve derivatives
of the smoothing lengths, is of the usual symmetrised form (Hernquist \&
 Katz 1989). The second part, which involves derivatives of the smoothing
lengths, arises because of the spatial and temporal variability of
 the $h_i$s. We shall
 refer to terms of this type as $\nabla h$ terms.
In order to evaluate the second part, we need to specify the way that
the smoothing
length
depends on the particle coordinates.
\noindent In order that the system conserve linear and angular momentum
as well as energy, we take
the $h_k$  to be functions only of the absolute distances between
particles. Here, we shall adopt the prescription given in Nelson
\& Papaloizou (1993), namely
 \be h_k = {\cal G}_k\left( \sum_{n=1}^{N} \chi_{kn} H_{kn}(| {\bf r}_k -
 {\bf r}_n |) \right) \label{hk} \ee
 where ${\cal G}$ and $H$ are arbitrary functions, and $\chi_{kn}$
 are arbitrary numbers, being zero when $k=n$.
 After differentiation of equation
 (\ref{hk}) with respect to ${\bf r}_i$, we obtain

 \be{{\partial h_{k}}\over{\partial{\bf r}_{i}}}
 =\delta_{ki}{\cal G}_k'\sum_{n=1}^{N}\left(\chi_{kn}H_{kn}'{{\bf r}_{kn}
 \over |{\bf r}_{kn}|}\right)+{\cal G}_k'\chi_{ki}H'_{ki}
 {{\bf r}_{ik}
 \over |{\bf r}_{ik}|},\label{dhkdri}\ee
 where a prime denotes differentiation of a function with respect to its
 argument. By combining equations (\ref{gradp3}) and (\ref{dhkdri}) we
 eventually obtain, for the
 pressure force on particle $i$, the expression

 \begin{eqnarray}
 {\bf F}_{P,i} & = & - \frac{1}{2} \sum_{j=1}^{N} m_i m_j
 \left( \frac{P_i}{\rho_{i}^{2}} +
 \frac{P_j}{\rho_{j}^{2}} \right) \left[
 \left.\frac{\partial W({\bf r}_{ij},h_i)}{\partial {\bf r}_i}
 \right|_{h_i const} +
 \left.\frac{\partial W({\bf r}_{ij},h_j)}{\partial {\bf r}_i}
  \right|_{h_j const}
  \right] \nonumber \\
 & & -\frac{1}{2} \sum_{j=1}^{N} m_i m_j \left( \frac{P_i}{\rho_{i}^{2}}
 + \frac{P_j}{\rho_{j}^{2}} \right)
 \frac{\partial W({\bf r}_{ij},h_i)}{\partial h_i} {\cal G}_{i}' \sum_{n=1}^{N}
 \chi_{in} H_{in}' \frac{{\bf r}_{in}}{| {\bf r}_{in} |} \nonumber \\
 & & -\frac{1}{2}\sum_{k=1}^{N} m_k \frac{P_k}{\rho_{k}^{2}}
 {\cal G}_k' \chi_{ki} H_{ki}'
 \frac{{\bf r}_{ik}}{| {\bf r}_{ik} |} \sum_{j=1}^{N} m_j
 \frac{\partial W({\bf r}_{kj},h_k)}{\partial h_k} \nonumber \\
 & & -\frac{1}{2}\sum_{k=1}^{N} m_k {\cal G}_k' \chi_{ki} H_{ki}'
 \frac{{\bf r}_{ik}}{| {\bf r}_{ik} |} \sum_{j=1}^{N} m_j
 \frac{P_j}{\rho_{j}^{2}} \frac{\partial W({\bf r}_{kj},h_k)}{\partial h_k}.
 \label{gradp4}  \end{eqnarray}

\noindent  This form of the pressure force leads to a fully
 conservative set of equations because it has been derived
from either of the potential functions (\ref{poten}) or (\ref{poten1}) which
are
 explicit functions of the inter-particle
 distances only.
\subsubsection {The addition of viscosity.}
If  dissipative mechanisms are included, then we no longer have a system that
is
 Hamiltonian. However, it is possible to ensure that the net rate
 of energy input into the system is equal to the time rate of change of
the
 Hamiltonian. The equations derived above for  inviscid systems
are thus modified so that the energy dissipated through viscosity goes into
thermal energy
in a consistent manner.
  For calculations in which artificial viscosity is included, we add an
artificial
 viscous pressure term, $\Pi_{ij}$, (Monaghan \& Gingold 1983),
in only the leading term of equation
 (\ref{gradp4}), not involving $\nabla h$ terms,
such that
$$\frac{P_i}{\rho_{i}^{2}} + \frac{P_j}{\rho_{j}^{2}}\rightarrow
\frac{P_i}{\rho_{i}^{2}} + \frac{P_j}{\rho_{j}^{2}} +\Pi_{ij}.$$
With this addition
$${\bf F}_{P,i}\rightarrow {\bf F}_{P,i} + {\bf{S}}_{visc, i}.$$
Where it has been used, we
have adopted the artificial viscous pressure given by Monaghan \& Gingold
(1983)
in the form
 \be \Pi_{ij} = \frac{1}{{\bar\rho}_{ij}}\left( - \alpha \mu_{ij} {\bar c}_{ij}
  + \beta \mu^{2}_{ij} \right) \label{Pi} \ee
 where the notation ${\bar A}_{ij} = \frac{1}{2}(A_{i}+A_{j}) $ has been used,
 c is the adiabatic sound speed, and
  \be  \mu_{ij} = \left\{ \begin{array}{cl}
  \displaystyle
   {\bar h_{ij}}\frac{{\bf v}_{ij}.{\bf r}_{ij}}{r_{ij}^{2}
  + \eta^{2}}   & \mbox{if ${\bf v}_{ij}.{\bf r}_{ij} < 0 $} \\
  0 & \mbox{otherwise.} \label{muij}
  \end{array}
  \right. \ee
  Here, the notation ${\bf v}_{ij}=({\bf v}_{i}-{\bf v}_{j})$ has been used,
and
  $\eta^{2}=0.01{\bar h}^{2}_{ij}$ prevents the denominator from vanishing.
\subsubsection{The entropy equation.}
For a representation of the entropy equation (\ref{dSdt}), we write,
 for particle $i$,
\be \left.{\partial {\cal U}_{i}\over \partial S_{i}}\right|_{\rho_{i}} \frac{d
S_{i}}{dt} =  {\cal H}_{i}.
\label{dSdti} \ee
This may be incorporated into a conservative scheme (see below).

\subsubsection{Conservation of energy.}
In the case of either a strictly adiabatic or barotropic inviscid flow,
the total energy, $E$ is conserved.
This may be written in the form
$$E=T_K+U+\Omega.$$
Here
$$T_K={1\over 2}\sum_{i=1}^{N}m_i|{\bf v}_i|^2$$
is the kinetic energy, $U$ is the potential function representing the thermal
energy
which we used above to derive the pressure force, and $\Omega$ represents the
gravitational
energy. Provided that $\Omega$ is
expressed as a function only of the distances between
particles,
it may be used to derive the gravitational force.

\subsubsection{The effect of viscosity.}
When viscous terms are included in the equations of motion, and we allow the
entropy
$S_i$ associated with particle $i$ to vary with time, we may derive an
equation for $E$ by taking the scalar product of ${\bf v}_i$ with the equation
of motion (\ref{dvdti}) and summing over $i$, so obtaining

\begin{eqnarray}
{dE\over dt}= \sum_{i=1}^Nm_i{\partial {\cal U}_i  \over\partial S_i}{d
S_i\over
dt}&-&\frac{1}{4}\sum_{i=1}^N \sum_{j=1}^{N} m_i m_j\Pi_{ij}
{{\bf v}_{ij}\cdot{\bf r}_{ij}\over |{\bf r}_{ij}|} \times  \nonumber \\
 & & \left[
 \left.\frac{\partial W({\bf r}_{ij},h_i)}{\partial |{\bf r}|_{ij}}
 \right|_{h_i const} +
 \left.\frac{\partial W({\bf r}_{ij},h_j)}{\partial |{\bf r}|_{ij}}
  \right|_{h_j const}
  \right] \label{Eeq}
 \end{eqnarray}
Here, the second term, involving the double summation, represents the viscous
dissipation,
which is necessarily positive definite.

\noindent If we write equation (\ref{dSdti}) in the form

\begin{equation}
{\partial {\cal U}_i \over\partial S_i}{d S_i\over dt}= {\cal H}_i
=\frac{1}{4} \sum_{j=1}^{N} m_j\Pi_{ij}
{{\bf v}_{ij}\cdot{\bf r}_{ij}\over |{\bf r}_{ij}|}
 \left[
 \left.\frac{\partial W({\bf r}_{ij},h_i)}{\partial |{\bf r}|_{ij}}
 \right|_{h_i const} +
 \left.\frac{\partial W({\bf r}_{ij},h_j)}{\partial |{\bf r}|_{ij}}
  \right|_{h_j const}
  \right] + {\cal L}_i ,\label{Eent}\end{equation}
where the first term on the right hand side represents
the heating rate per unit mass due to viscosity and ${\cal L}_i$ represents the
net non-viscous
heating rate per unit mass, then we see that equation (\ref{Eeq}) becomes
$${dE\over dt}= \sum_{j=1}^Nm_j{\cal L}_j.$$
Thus the rate of increase of the total energy is equal to the net non viscous
heating rate
as required.

{\noindent} The evolution of the entropic function, $K$, defined in Section
(2),
 is then given by the expression

\be  \frac{dK(S_i)}{dt} = (\gamma -1){\cal H}_i \rho^{1-\gamma}_i \ee

\noindent Using the second law of thermodynamics
$$
{d {\cal U}_i\over dt}= {\partial {\cal U}_i \over\partial S_i}{d S_i\over dt}
+{P_i\over \rho^2_i}{d \rho_i\over dt},$$
we may derive an equation for the rate of change of the internal
energy per unit mass for particle $i.$
Using (\ref{Eent}) and differentiating equation (\ref{den2}), this may be
written in the form
 \begin{eqnarray}
{d {\cal U}_i\over dt} & = & {\cal H}_i +
 \frac{1}{2}\left( \frac{P_i}{\rho_{i}^{2}} \right) \sum_{j=1}^{N }m_j
 {{\bf v}_{ij}\cdot{\bf r}_{ij}\over |{\bf r}_{ij}|}
 \left[
 \left.\frac{\partial W({\bf r}_{ij},h_i)}{\partial |{\bf r}_{ij}|}
 \right|_{h_i const} +
 \left.\frac{\partial W({\bf r}_{ij},h_j)}{\partial |{\bf r}_{ij}|}
  \right|_{h_j const}
  \right] \nonumber \\
 & &
 +\frac{1}{2}\left( \frac{P_i}{\rho_{i}^{2}} \right)\sum_{j=1}^{N} m_j
 \frac{\partial W({\bf r}_{ij},h_i)}{\partial h_i} {\cal G}'_i \sum_{n=1}^{N}
 \chi_{in} H_{in}' \frac{{\bf v}_{in}\cdot{\bf r}_{in}}{| {\bf r}_{in} |}
\nonumber \\
 & &
 +\frac{1}{2}\left( \frac{P_i}{\rho_{i}^{2}} \right)\sum_{j=1}^{N} m_j
 \frac{\partial W({\bf r}_{ij},h_j)}{\partial h_j} {\cal G}'_j \sum_{n=1}^{N}
 \chi_{jn} H_{jn}' \frac{{\bf v}_{jn}\cdot{\bf r}_{jn}}{| {\bf r}_{jn} |}.
 \label{uintequ}
\end{eqnarray}
Equation (\ref{uintequ}) may thus be used as an {\em equivalent}
alternative to the entropy equation (\ref{dSdti}).

\vspace{2mm}

\noindent  At this point, we note that the standard method of deriving the SPH
 equations of motion - i.e. by multiplying the continuum equations by $W$,
 integrating over the solution domain, and integrating the
 resulting expressions
 by parts (see Hernquist \& Katz 1989) - will not result in conservative
equations of motion when
spatially
 variable smoothing lengths are employed, even if the $\nabla h$ terms that
 arise from such an approach are included. The reason for this is that, in the
standard
 method, the pressure forces experienced by the particles
 can be expressed in the form of a sum over pairwise interactions.
 However, when the
 smoothing lengths are expressed in terms of inter-particle distances, it is
 apparent that the pressure force on a particle $i$ can no longer be
 expressed in the form
 of a sum of simple pairwise interactions, but simultaneously involves
 communication
 with other particles
 in the system - namely all the particles which contribute to $h_i$, and also
 those
 particles in the system that receive a contribution from particle $i$ to
 the calculation of their own smoothing lengths.

 {\noindent} In principle, it is not necessary
 to employ either of the symmetrisation
 procedures described by equations (\ref{hij}) or (\ref{Wij})
 when the equations
 of motion are derived from a Hamiltonian. When the Hamiltonian is formed
 using either of the so-called gather or scatter formulations (for definition
 of these terms see
 Hernquist \& Katz 1989), the resulting equations of motion are
 symmetric and will conserve
 both momentum and energy. However, the form in which they appear does not
 allow the artificial viscosity in equation (\ref{Pi}) to be
 incorporated in the usual manner.

\subsubsection{The smoothing kernel.}
  The smoothing kernel used is that proposed by Monaghan \&
 Lattanzio (1985),
  and takes the form
 \be W(u,h) = \frac{1}{\pi h^{3}} \left\{ \begin{array}{cl}
  \displaystyle
  1-\frac{3}{2}\left(\frac{u}{h}\right)^{2} + \frac{3}{4}\left(\frac{u}{h}
  \right)^{3}   & \mbox{$0\le u/h \le 1$} \\
  \displaystyle
  \frac{1}{4}\left[ 2-\left(\frac{u}{h}\right)\right]^{3}
  & \mbox{$ 1\le u/h \le 2 $} \\
  \displaystyle
  0   & \mbox{$ u/h \ge 2 $}
  \end{array}
  \right.  \label{kernel} \ee
  where $u = |{\bf r}_i - {\bf r}_j| .$

 \subsubsection{Smoothing lengths.}
 There are two basic requirements that must be met by the method chosen
 to calculate the individual particle smoothing lengths. Firstly,
 the smoothing lengths must be explicit functions of the distances between
 particles,
 and secondly, the number of particles with which a given particle interacts
 should be roughly constant. We use a method similar to that suggested
 by Hernquist \& Katz (1989), in order to keep the number of neighbours
 within $2h_i, {\cal N}_{TOL}$, constant. Having obtained this list of
 nearest neighbours, $h_i$ is then defined to be

 \be  h_i = \frac{1}{2} | {\bf r}_i - {\bf r}_{imax} | \label{hi} \ee
 where ${\bf r}_{imax}$ is the position vector of particle $i$'s most
 distant nearest neighbour. In terms of our general functional form for
 $h_i$ given by equation (\ref{hk}), the functions ${\cal G}(x)=x$
 $\chi_{in}=\delta_{n,imax}$ , and
 $H(x)=\frac{1}{2} x$.
 From equation (\ref{gradp4}), the pressure force on particle $i$ then becomes

 \begin{eqnarray}
 {\bf F}_{P,i} & = & - \frac{1}{2} \sum_{j=1}^{N} m_i m_j
 \left( \frac{P_i}{\rho_{i}^{2}} +
 \frac{P_j}{\rho_{j}^{2}} \right) \left[
 \left.\frac{\partial W({\bf r}_{ij},h_i)}{\partial {\bf r}_i}
 \right|_{h_i const} +
 \left.\frac{\partial W({\bf r}_{ij},h_j)}{\partial {\bf r}_i}
  \right|_{h_j const}
  \right] \nonumber \\
 & & -\frac{1}{4} \sum_{j=1}^{N} m_i m_j \left( \frac{P_i}{\rho_{i}^{2}}
 + \frac{P_j}{\rho_{j}^{2}} \right)
 \frac{\partial W({\bf r}_{ij},h_i)}{\partial h_i}
   \frac{{\bf r}_{i\  imax}}{| {\bf r}_{i\  imax} |} \nonumber \\
 & & -\frac{1}{4}\sum_{k=1}^{N} m_k \frac{P_k}{\rho_{k}^{2}}\delta_{i\  kmax}
 \frac{{\bf r}_{kmax\  k}}{| {\bf r}_{kmax\  k} |} \sum_{j=1}^{N} m_j
 \frac{\partial W({\bf r}_{kj},h_k)}{\partial h_k} \nonumber \\
 & & -\frac{1}{4}\sum_{k=1}^{N} m_k \delta_{i\  kmax}
 \frac{{\bf r}_{kmax\  k}}{| {\bf r}_{kmax\  k} |} \sum_{j=1}^{N} m_j
 \frac{P_j}{\rho_{j}^{2}} \frac{\partial W({\bf r}_{kj},h_k)}{\partial h_k}
 \label{gradp5}
 \end{eqnarray}
 and similarly for the equation governing the evolution of the specific
 thermal energy we obtain from (\ref{uintequ}) the expression

 \begin{eqnarray}
{d {\cal U}_i\over dt} & = & {\cal H}_i +
 \frac{1}{2}\left( \frac{P_i}{\rho_{i}^{2}} \right) \sum_{j=1}^{N }m_j
 {{\bf v}_{ij}\cdot{\bf r}_{ij}\over |{\bf r}_{ij}|}
 \left[
 \left.\frac{\partial W({\bf r}_{ij},h_i)}{\partial |{\bf r}_{ij}|}
 \right|_{h_i const} +
 \left.\frac{\partial W({\bf r}_{ij},h_j)}{\partial |{\bf r}_{ij}|}
  \right|_{h_j const}
  \right] \nonumber \\
 & &
 +\frac{1}{4}\left( \frac{P_i}{\rho_{i}^{2}} \right)\sum_{j=1}^{N} m_j
 \frac{\partial W({\bf r}_{ij},h_i)}{\partial h_i}
   \frac{{\bf v}_{i\  imax}\cdot{\bf r}_{i\  imax}}{| {\bf r}_{i\  imax} |}
\nonumber \\
 & & +\frac{1}{4}\left( \frac{P_i}{\rho_{i}^{2}} \right)\sum_{j=1}^{N}  m_j
 \frac{{\bf r}_{j\  jmax}\cdot {\bf v}_{j\  jmax}}{| {\bf r}_{j\  jmax} |}
 \frac{\partial W({\bf r}_{ij},h_j)}{\partial h_j}\label{uinteq1}
 \end{eqnarray}
 We note that, in order for the potential function given by equation
 (\ref{poten}) to
  be a  continuous function of its arguments,
 the functional form adopted for the
 $h_i$s must  be  continuous. Consequently, the number of
 nearest neighbours, ${\cal N}_{TOL}$, that determines the value of each
 $h_i$ should be {\it strictly} constant throughout the calculation,
 and particle
 $imax$ in equation (\ref{hi}) must always be
 the `${\cal N}_{TOL}$th' nearest
 neighbour. This point is discussed further in Section (5).

 We also note that the functional form (\ref{hi})  used to calculate
 the smoothing lengths
 introduces additional statistical fluctuations into the pressure forces.
 From the momentum equation (\ref{gradp5}) it is apparent
 that when particle $i$'s
 nearest neighbour, $imax$, changes its identity as the system evolves in
 time, fluctuations will arise in the pressure forces experienced by
 particle $i$. Technically speaking, the potential function defined by
 equation (\ref{poten}) is itself continuous, but it  has a discontinuity in
its
 first derivative, the magnitude of which is determined by the
 $\nabla h $ terms. An alternative method for calculating
 the smoothing lengths, which reduces these fluctuations, is to first find
 the ${\cal N}_{TOL}$ nearest neighbours and to then calculate $h_i$ from
 this list of nearest neighbours according to

 \be h_i = \frac{1}{N_{far}} \sum_{n=1}^{N_{far}} \frac{1}{2} | {\bf r}_i
 - {\bf r}_n | \label{hn6} \ee
 where the summation is over particle $i$'s  $N_{far}$ most distant
 nearest neighbours. In particular, we have
 experimented with using $N_{far}=6$
 so that the value of $2h_i$ then becomes equal to the mean
 distance to particle $i$'s six most distant nearest
 neighbours. In terms of the general functional form for $h_i$ given by
 equation (\ref{hk}), the functions ${\cal G}(x)=x, \chi=\delta_{i \; ifar},
 H(x)=x/2N_{far}$, where each $i_{far}$ represents one of particle $i$'s
 $N_{far}$ most distant nearest neighbours.

 \subsubsection{Gravitational forces.}
 An implementation of the Barnes-Hut hierarchical tree algorithm
 (Barnes \& Hut 1986; Hernquist 1987) was used in the calculations for
 which the computation of the force ${\bf F}_{G,i},$ due to self-gravity was
required.
 The interparticle potential was softened by the method of spline softening
 (Gingold \& Monaghan 1977), with a constant value of the softening length
 used throughout. When calculating the gravitational force on a given
 particle $i$, an opening angle of $\theta=0.6$ (where $\theta=s/d; s=$size of
 cell, $d$=distance to centre of mass of cell),
 was used for cells whose centre of mass lay
 outside of the softening radius, with forces being calculated up to
 quadrupole order. For cells lying within the softening radius, an opening
 angle of $\theta = 0.3$ was employed, and forces were computed up to
 monopole order only.

 This method of
 calculation gives an approximation to
the gravitational
potential in such a way that it is no longer strictly a sum of contributions
dependent only on the distances between particles, and this will lead to some
small $(\Delta E \ls 1 \%)$ error in
the total energy conservation.

 \subsubsection{Time integration.}
 The standard second order leap-frog scheme, with the modification
 proposed by Hernquist \& Katz (1989) for estimating time-centred
 velocities in the viscous pressure term (\ref{Pi}),
 the energy equation (\ref{uintequ}),
 and the entropy equation (\ref{Eent}),
 was used. The time step was
 determined by the expression

 \be \delta t = {\cal Q}\, \mbox{Min}_{i} \frac{h_{i}}{
  c_{i} +1.2(\alpha c_{i} + \beta\, \mbox{{max}}_{j} |\mu_{ij}|)} \label{time}
 \ee
 where the factor ${\cal Q}$ is a numerical constant usually taken to be in
 the range ${\cal Q} \approx 0.1 - 0.3.$

 \section{Numerical tests.}

 A number of test calculations have been performed using different versions
 of our SPH code in order to compare their ability to reproduce
 the solutions for problems that have either been
 studied using independent numerical
 techniques, or which have known analytical solutions.
 The main purpose of this paper, however, is to study the
 effect of including the $\nabla h$ terms on the conservation of energy, and
 the emphasis will be on those calculations designed to test
 conservation properties.
 Tests have also been performed to explore the
 possibility that the inclusion of $\nabla h$ terms in the manner
 described above may lead to increased numerical diffusivity.

 \subsection{Shock-tube calculations.}

 Extensive one-dimensional Riemann shock-tube calculations (see Sod 1978;
 Monaghan \&
 Gingold 1983), have been performed.
 The initial conditions were:

 \be
 \begin{array}{llll}

 \rho  =  1, &   P  =  1, &   v  =  0, & {\rm for} \mbox{ } x  <  0;  \\
 \rho  =  0.25, &  P  =  0.1795, &  v =  0, & {\rm for} \mbox{ }  x  \ge  0.

 \end{array}
 \label{shcond}
 \ee
Here $x$ is a cartesian coordinate and the calculations were for an ideal gas
with specific heat ratio $\gamma=1.4$. The equations integrated were the
equation of motion (\ref{dvdti}) using (\ref{gradp5}) and the entropy equation
(\ref{Eent}) with ${\cal L}_i = 0.$
 Plots of the density, pressure, velocity field, and entropic function
 $K$, are shown in Figs. (1.a - 1.d)
 for the case when the $\nabla h$ terms
 were neglected, and in Figs. (2.a - 2.d) for the
 case when the $\nabla h$ terms
 were included. In both cases shown, 400 equal mass particles were distributed
 in the range $ -0.6 \le x \le 0.6 $  so as to satisfy equation (\ref{shcond}).
 The smoothing lengths were allowed to vary in both space and time, with
 the number of nearest neighbours being fixed at ${\cal N}_{TOL}=4$, using
 a renormalised, one-dimensional version of the smoothing
 kernel (\ref{kernel}). This
 corresponds to $\approx 32$ neighbours for a spherical kernel
 in three-dimensions. For the calculations presented here, the
 artificial viscosity parameters in equation (\ref{Pi})
 were chosen to be $\alpha=1., \beta=1., \eta^2 = 0.01$, and the time-step
 was calculated using equation (\ref{time}) with ${\cal Q}=0.2$.

 The results shown in both Figs. (1.) and (2.) are almost identical in form,
 and show a satisfactory agreement with the analytical solution. The shock
 front located between $x=0.2 - 0.25$ is broadened over a range in $x$ of
 $ \approx 2h_i$, with the contact discontinuity, positioned at
 $x \approx 0.1 $ having a similar width. The slight overshoot in the density
 and pressure at $x \approx -0.2 $ appears to be due to the fact that
 the values of the smoothing lengths become smaller at this point in order
 to satisfy the constraint that the number of nearest neighbours be constant.
 This effect was seen to decrease when the number of nearest neighbours
 ${\cal N}_{TOL}$ was increased, though this also had the effect of broadening
 the shock front.

 \subsection{Gravitational collapse of cold gas sphere.}

 The adiabatic collapse of an initially isothermal gas sphere has been
 calculated using different versions of our code in order to test the
 qualitative effects of including the $\nabla h$ terms in a three-dimensional
 calculation. The initial conditions were chosen to match those of
 Evrard (1988) and Hernquist \& Katz (1989) so as to facilitate comparison
 with their P3MSPH and TREESPH codes respectively, and with a finite-difference
 code (Thomas 1987).

 \noindent Following Evrard (1988), the initial state is an ideal
 gas sphere of radius
 $R$ and mass $M_T$, with a density profile given by
 \be \rho = \frac{M_T}{2\pi R^2} \frac{1}{r}, \label{rho} \ee
where $r$ is the usual spherical polar radius. The initial conditions were
set up by placing the particles on a uniform grid which was then stretched
 to achieve the required density distribution.
 Initially, the gas sphere was isothermal, with specific thermal energy
 $u = 0.05G M_T/R$. The ratio of the specific heats is $\gamma = 5/3.$

\noindent As in Evrard (1988), the scales used in presenting the results are:
density, $\rho_* = 3M_T/4\pi R^3$; energy, $ u_* = GM_T/R$;
 pressure, $ P_* = \rho_* u_*$; and velocity, $v_* = (GM_T/R)^{1/2}.$ Time
 is shown in units of the free-fall time at the initial outer radius,
 $ t_* = (\pi^2/8G)^{1/2} R^{3/2}M_T^{-1/2} $.
The equations integrated were the
equation of motion (\ref{dvdti}) using (\ref{gradp5}), and the entropy equation
(\ref{Eent}) with ${\cal L}_i=0.$
The results for the
 calculation in which the $\nabla h$ terms were neglected are shown
 in Figs. (3. - 6.), and those for which the $\nabla h$ terms were included
 are shown in Figs. (7. - 10.).
 The gas sphere begins to collapse due to the low value of the internal energy,
 and as it does so the ratio of the thermal to the gravitational energy
 increases since
 $ \gamma > 4/3 $.  At later times, a central bounce occurs and a shock
 wave propagates outwards through the collapsing sphere, with the system
 converting most of its kinetic energy into thermal energy between times
 $ t \approx 0.8t_* $ and $ t \approx 1.2t_* $. Eventually, the sphere
 tends towards an approximate equilibrium configuration with
 $U \approx - \Omega/2 $.

 Both of the calculations displayed in Figs. (3. - 6.) and Figs. (7. -
 10.)
 show excellent agreement with the SPH calculations of Evrard (1988),
 and those of
 Hernquist \& Katz (1989). Particularly encouraging is the good agreement
 found between our calculations and the results obtained by Evrard (1988)
 using a one-dimensional finite difference code containing 250 zones
 (Thomas 1987). The strength and position of the shock at time
 $ t=0.8t_* $ is well represented by our calculations, and the thermal
 energy profile behind the shock shows the slow rise with radius displayed
 by the one-dimensional model. Similarly, the velocity profiles from our
 calculations, shown in Fig. (6.) and Fig. (10.), indicate
 good general agreement with those of the one-dimensional calculation, though
 there is a larger scatter in these results than in those for the density and
 pressure caused largely by non-synchronous evolution of the system at
 different angular positions around the sphere.

 When comparing the results of our two SPH
 calculations, small differences may be observed. Firstly, the results
 for the case in which the $\nabla h$ terms were included appear to suffer
 from a slightly larger scatter than the results from the calculation in
 which the $\nabla h$ terms were neglected. This `noise' appears to be due
 to the functional form, (\ref{hi}), adopted for calculating the $h_i$s, which,
 as described in Section (3.1.10), introduces increased statistical
fluctuations
 into the pressure forces.
 However, the overall effect of these increased fluctuations appears to be
small.
 Closer inspection of the results shows that including the
 $\nabla h$ terms actually leads to improvements in the performance
of the code due to energy being conserved more
 accurately.
 For example, at time $t=1.2t_*$, the sharp peak and drop in the thermal energy
 located at $r \approx 0.6R $ in Evrard's one-dimensional calculation is more
 apparent in our results if $\nabla h$ terms are included
 than if they are neglected. Consequently, the corresponding feature in the
 pressure profile is also more accurately represented. Energy was conserved
 to a level of $\approx 1.0 \%$ when $\nabla h$ terms were included, and
 to a level of $\approx 4.3 \%$ when neglected.

 The calculation was repeated using the energy equation (\ref{uinteq1}) with
 the $\nabla h$ terms being both included and  neglected, respectively.
 As expected, the qualitative outcome of these calculations were almost
 identical to their equivalent calculations displayed in Figs. (3. - 10.)
 in which the entropy equation
 (\ref{Eent}) was integrated. If the $\nabla h$ terms are neglected, then no
 apparent qualitative improvement in the results is obtained
  by using the thermal energy equation
 (\ref{uinteq1}) -
 which leads to conservation of total energy but {\em not} entropy, instead
 of the entropy equation (\ref{Eent}) -  which leads to the conservation of
 entropy but not energy.

\noindent The calculation was also performed using the functional form for
 the $h_i$s given by equation (\ref{hn6}) with $N_{far}=6$, and ${\cal
N}_{TOL}$
 taking the value $50 \pm 1$.
 In this case
 the $\nabla h$ terms were included, and the entropy equation (\ref{Eent})
 was integrated.
 The results were
 almost identical to those shown in Figs. (7.) - (10.), except that the level
 of scatter was reduced slightly.
 Energy was conserved to a level of $\approx 0.6 \%$.

\noindent Obviously,
 the inclusion of $\nabla h$ terms does not alter the ability of SPH
 to reproduce the salient features present in the problems studied thus far,
 and can in fact improve the qualitative outcome of the calculations.

 \section{Conservation of energy.}

 Two different sets of test calculations were performed in order to
 test the energy conservation properties of different versions of our code.
 These were collision calculations between identical polytropes,
 including the effects of self-gravity and artificial viscosity,
 and the free expansion of non self-gravitating polytropes in which
 the effects of viscosity were ignored.

 \subsection{Colliding polytropes.}

 Numerical calculations of collisions between identical $n=3/2$ polytropes
 have been performed. Equilibrium polytropes, each consisting of $N=2085$
 particles, were set-up by evolving an initially uniform gaseous sphere
 in the presence of damping until a stationary state was achieved (Lucy 1977;
 Goodman \& Hernquist 1991). Rather than being allowed to collide after
free falling from
 a distance, the two identical polytropes were placed adjacent to one another
 with their surfaces just in contact. The polytropes were then given
 the free-fall velocities that they would have if they had started
 from a state of rest, with an initial
 separation between
 their centres equal to six times their diameter. This procedure was followed
 in order to reduce the computational costs.

 \noindent Collision calculations for six different versions of our code were
performed,
 the details of which are summarised in Table 1.
In these calculations the equations integrated were the
equation of motion (\ref{dvdti}) using (\ref{gradp5}), and either the entropy
equation (\ref{Eent}) or the internal energy equation (\ref{uinteq1}).
In all cases ${\cal L}_i=0.$
 A typical time sequence
 is shown in Fig. (11.), and the evolution of the total energy for
 a number of the calculations described in Table 1. are
 shown in Fig. (12.). For all the calculations presented,
 a value of ${\cal Q}=0.2$ was used in equation (\ref{time})
 to determine the time
 step, and the artificial viscosity parameters in equation (\ref{Pi}) were
 $\alpha=0.5, \beta=1$.

\noindent  As the polytropes collide, a shock forms at the interface converting
 most of the kinetic energy into thermal energy. A period of rapid re-expansion
 then occurs due to the presence of large pressure gradients in the fluid,
 but because the system remains bound, the resulting single polytrope undergoes
quadrupole
 oscillations which damp out as the system tends towards an equilibrium state.

\subsubsection{$\nabla h$ terms neglected.}
 From the results for the evolution of the total energy shown in Fig. (12.),
 it
 is obvious that large differences exist between the results for the
 two calculations in which the $\nabla h$ terms were neglected. It appears
 that the total energy is conserved with reasonable accuracy (at the $\approx
 2 \%$ level) if the energy equation (\ref{uinteq1})
 is integrated, whereas errors at
 the level of $\approx 10\%$ are incurred if the entropy equation
 (\ref{Eent}) is
 integrated. These errors largely occur between the times $t=0.1$ and
 $t=0.3$ during the initial compression and rapid re-expansion phase.
 However, the apparent improvement in the conservation of energy
 obtained by integrating the energy equation is somewhat illusory since
 entropy is no longer conserved (Hernquist 1993). This is
 because neglecting the $\nabla h$ terms in both (\ref{gradp5}) and
(\ref{uinteq1}) is consistent with conservation of total energy but
is inconsistent with the correct equation for the entropy.
 We have performed further
 calculations that show that the errors incurred in the conservation
of the total energy
when integrating
 equation (\ref{uinteq1}) are largely due to the use of velocities that are not
 properly time-centred and  from the
 use of a tree-code to compute the self-gravity of the fluid. Also, calculation
 $C4$ in Table 1. was repeated using a value for ${\cal Q}$ in equation
 (\ref{time})
 of ${\cal Q} = 0.03$. Violations in energy conservation still occurred at the
 level of $9.75 \%$, implying that non-conservation is not induced by
 time-stepping errors, but is instead due to a physical change of the energy
 in the system introduced by the time variation of the smoothing lengths.

 \subsubsection{$\nabla h$ terms included.}

  It is apparent from Fig. (12.) that the inclusion of $\nabla h$ terms
 results in a dramatic improvement in the conservation of total
 energy. The entropy equation (\ref{Eent}) was integrated for all runs
 displayed in Fig. (12.) in which the $\nabla h$ terms were included.
 As mentioned in Section (3.1.10),
 by allowing the value of ${\cal N}_{TOL}$ to vary
 from a single, fixed value, temporal discontinuities are
 introduced into the potential energy of the system, and energy is no longer
 exactly conserved. This point is illustrated by our results, which show
 the conservation of total energy slowly degrade as the
 value of ${\cal N}_{TOL}$ is
 allowed to vary from ${\cal N}_{TOL}=50$ to ${\cal N}_{TOL}=50 \pm 1$, and
 ${\cal N}_{TOL}=50 \pm 2$ respectively. However,
 even with ${\cal N}_{TOL}$ varying
 by $\pm 2$, the errors incurred are still relatively small $(\approx 3 \%)$,
 and for less violent scenarios are almost negligible.
 Computational advantage is gained by allowing ${\cal N}_{TOL}$ to
 vary slightly about some fixed value rather than maintaining constancy.
 Any method used to estimate the $h_i$s at the beginning of a time step
 has a greater chance of satisfying the demand that the number of
 nearest neighbours lies within a small range about some value, rather
 than being equal to a single, specific value. Therefore, the estimated
 $h_i$s have to be adjusted less often and the execution speed of the code
 is increased.
 Errors incurred in the conservation of energy for the
 run in which ${\cal N}_{TOL}$ took a single value (i.e. ${\cal N}_{TOL}=50$),
 were due to residual discretisation errors, and the use of a tree-code to
 compute gravitational forces.

 The collision calculation was also performed by including the $\nabla h$ terms
 and integrating the thermal energy equation (\ref{uinteq1}). Although we
 do not display the results from this calculation in Fig. (12.), because
 integrating either of the equations (\ref{Eent}) or (\ref{uinteq1})
 is formally equivalent when $\nabla h$ terms are included, we found
 that the evolution of the total energy was very similar to the case
 in which equation (\ref{uinteq1}) was integrated and the $\nabla h$ terms
 were neglected, as should be expected.
 The deviation from perfect energy
 conservation in this case is again due to the
 use of velocities that were not properly time-centred
 in equation (\ref{uinteq1}),
 and the use of a tree-code.

 \subsubsection{Timing tests.}

 The CPU time per time step was monitored for a number of collision
calculations
 in order to estimate the increase in computation time arising from the
 inclusion of
 the $\nabla h$ terms. The results are presented in Table 2. The two parameters
 that were varied were the number of nearest neighbours within $2h_i$,
 ${\cal N}_{TOL}$, and the functional form used to calculate the smoothing
 lengths, which was either equation (\ref{hi}) or equation (\ref{hn6}). A total
 of 4170 particles were used in each of the calculations.
 As mentioned in Section (5.1.2), allowing the value of ${\cal N}_{TOL}$ to
 vary about some value, 50 say, leads to a more efficient
 code, since less time is spent trying
 to iterate $2h_i$ towards a value such
 that it exactly encloses the
 $50^{th}$ nearest neighbour. It is apparent from Table 3 that the inclusion of
 the extra $\nabla h$ terms leads to only a small $(\ls 10 \%)$ increase
 in the computational
 overhead when equation (\ref{hi}) is used to calculate the smoothing lengths.
 When equation (\ref{hn6}) is used, with $N_{far}=6$,
 then the computational expense is greater,
 due to it being necessary to find the positions of the six most distant
 nearest neighbours from which the $h_i$s are then calculated.
 However, a rather naive sorting algorithm was used to perform this task,
 so we expect to be able to decrease the computational requirements with a
 more sophisticated approach in future calculations. All calculations shown
 in Table 3 were performed on a SUN IPX workstation,
 operating at 33MHz with 16 Mbytes of core storage.

 \subsection{Expanding polytropes.}
 In order to  test the
 different forms of the SPH algorithm under less extreme conditions,
 and in the absence of gravitational and viscous effects, a series
 of test calculations were performed on the free expansion of
 gaseous polytropes.

\noindent  An $n=3/2$ equilibrium polytrope, consisting of $N=4945$ particles,
was set
 up in the manner previously described in Section (5.1).
 At time $t=0$, the self-gravity
 was switched off. As expected, the absence of pressure at the surface boundary
 causes a rarefaction wave to propagate through the polytrope which
 expands freely into space, with the thermal energy being converted
 into the kinetic energy of the fluid's bulk motion. Artificial
 viscosity was neglected, though relation (\ref{time})
 was used to calculate the time step, with
 $\alpha=0.5, \beta=1., {\cal Q}=0.05$.

\noindent A number of free-expansion calculations were performed, with the run
 parameters being summarised in Table 3. The
 variation of the total energy in the system as a function of time
 is plotted in Fig. (13.) for a number of the runs described in Table 3.

\noindent Although the free expansion of a non self-gravitating polytrope is a
 considerably
 less violent phenomenon than the collision between two identical polytropes,
 a similar trend is seen to occur in the results for the conservation of total
 energy, though the discrepancies are less marked.

 \subsubsection{$\nabla h$ terms neglected.}
 From Fig. (13.) it is obvious that once again the use of the energy equation
 (\ref{uinteq1}) rather than the entropy equation (\ref{Eent})
 results in energy being
 conserved to a higher degree of accuracy when $\nabla h$ terms are not
 included. If the
 energy equation is integrated then errors incurred in the conservation of
 the total energy are small, and mainly arise from the fact that the velocities
 used
 in equation (\ref{uinteq1}) are not correctly time-centred. In this case,
 however, it can be shown that entropy is not conserved.
 Errors at the level of $\approx 1.5 \% $
 arise if the entropy equation (\ref{Eent}) is used instead of (\ref{uinteq1}).
 This is because
 the total energy is not conserved
 if time varying smoothing lengths are used and the $\nabla h$ terms
 are neglected when
 modeling strictly adiabatic flows.
 Thus, errors of comparable
 magnitude are incurred in the
 conservation of different fundamental properties of the fluid when
 variable smoothing lengths are employed, depending on whether one chooses to
 integrate the entropy equation (\ref{Eent}) or
 the energy equation (\ref{uinteq1}). We note that the time period during which
 most of the energy was lost from the system in calculation $E3$ (see Table
3.),
 i.e. $t=0.0 - t=0.2$, corresponds to $\approx 1200$ time steps in our
 simulations, and represents the point when more than $99.5 \%$ of the initial
 thermal energy in the system has been converted in to kinetic energy -
 implying that non-conservation is not largely due to time stepping errors.

 \subsubsection{$\nabla h$ terms included.}
 It is apparent from Fig. (13.) that the inclusion of the $\nabla h$ terms
 results in much improved energy conservation.
 If the value of ${\cal N}_{TOL}$ is kept constant, and the entropy equation
 is used, then energy is conserved almost exactly, with the remaining deviation
 being due to residual discretisation errors. If ${\cal N}_{TOL}$ is
 allowed to vary slightly about a fixed value
 (i.e. ${\cal N}_{TOL} = 50 \pm 2$), then small errors ($\approx 0.15 \%$)
 occur, which are substantially smaller than the errors resulting from
 integrating the entropy equation and neglecting the $\nabla h$ terms.
 When equation (\ref{uinteq1}) is integrated instead of equation (\ref{Eent}),
 then small errors, the evolution of which
 are similar to those from calculation $E6$ in Table 3., are
 incurred (see run $E5$ in Table 3.).
 We do not present the results from this calculation in Fig. (13.)
 for similar
 reasons to those given in Section (5.1.2).

 \section{Statistical fluctuations and numerical diffusivity.}
 This last series of test calculations is concerned with the increased
 statistical fluctuations resulting from the inclusion of the
 $\nabla h$ terms, and in particular the functional form, (\ref{hi}),
  adopted for
 calculating the $h_i$s.
 In order to examine the importance of this effect, two test calculations
 were carried out.
 \subsection{Pressure force fluctuations.}
 Firstly, the pressure forces on a random sub-set of the particles were
 monitored throughout the free expansion of an $n=3/2$ polytrope.
 An illustrative selection from these results are presented, that
 represent the forces experienced by the particle
 as a function of time in the $x$, $y$, and
 $z$ directions respectively. Fig. (14.a) shows the forces experienced during
 a calculation for which the $\nabla h$ terms were neglected,
 with the entropy equation being integrated, and Fig. (14.b)
 shows the forces on the same particle during a calculation in which
 the $\nabla h$ terms were included and equation (\ref{hi}) was used to
 calculate the smoothing lengths. A similar calculation was also
 performed, with the smoothing lengths being calculated according
 to the relation (\ref{hn6}), and $N_{far}=6$.
 The results from this calculation
 are presented in Fig. (14.c)

 \noindent It is apparent from Fig. (14.b) that the use of equation (\ref{hi})
 does
 lead to increased statistical fluctuations in the pressure forces
 experienced by the particles. These fluctuations, however,
 are generally not large
 relative to the absolute value of the overall pressure forces,
 and are found to decrease for increasing numbers of particles.  The results
 shown in Fig. (14.c) show that it is possible to more or less remove these
 excess statistical fluctuations by simply increasing the number of
 particles contributing to the functional form used to calculate the
 smoothing lengths. The fact that the contribution of
 only a small number of particles $(N_{far}=6)$ is required
 to reduce the pressure force fluctuations adds weight to the argument that
 the $\nabla h$ terms should be regarded as essentially being correction terms.
 Up to now, however, we have not in general used equation (\ref{hn6})
 to calculate
 the smoothing lengths since the optimal method for implementing this
 relation has
 yet to be found. Finding the furthest $N_{far}$
 nearest neighbours for each particle is a non trivial computational
 problem when
 efficiency is an important consideration.
 \subsection{Numerical diffusion.}
 Calculations of oscillating $n=3/2$ polytropes were performed
 in order to examine whether the increased statistical fluctuations
 have the effect of increasing
 the numerical diffusivity of SPH.
 These used $N_{TOL}=50 \pm 1.$
 At time $t=0$, oscillations were initiated in an equilibrium polytrope,
 consisting of $N=4945$ particles, by imposing an homologous contraction.
 The results are presented in Fig. (15.) which, when moving from top to
 bottom in each panel, shows the time
 dependence of the thermal energy, kinetic energy,
 total energy, and gravitational potential energy. Fig. (15.a)
 shows the time sequence for a calculation in which the $\nabla h$ terms
 were included and equation (\ref{hi}) was used to calculate the smoothing
 lengths, and Fig. (15.b) displays the oscillations for a similar
 calculation in which the $\nabla h$ terms were neglected. In both of
 these calculations, a small amount $(\alpha=0.1, \beta=0.)$ of artificial
 viscosity was used in order to stabilise the time integration scheme,
 though  the
 dissipated kinetic energy was not converted into thermal energy. Instead, the
energy removed by
 the viscosity was simply allowed to leave the system, thus the polytropic
equation of state was adopted throughout.

 \noindent In both calculations, the period of oscillation is accurate to
within
 a few
 per cent of that expected from analytical considerations (see Cox 1980).
 Contrary to initial expectations, however, it is found that the oscillations
 damp out more quickly if the $\nabla h$ terms are not included in the
 calculations, since the improved energy conservation gained by including
 the $\nabla h$ terms more than off-sets any detrimental effects due to
 increased fluctuations in the pressure forces. If the amplitude of the
 oscillations is measured by the height of the thermal energy curves in
 Figs. (15.a) and (15.b), then the oscillations damp by $\approx 35 \%$ if the
 $\nabla h$ terms are included, and by $\approx 58 \%$ if the $\nabla h$ terms
 are ignored, after thirteen oscillation periods.

 \section{Discussion and conclusions.}

 A reformulation of the SPH equations of motion that contain terms accounting
 for the local variability of the smoothing lengths has been presented. In
 a previous paper (Nelson \& Papaloizou 1993) a set of
 conservative equations for a barotropic fluid were derived.
 In this current paper the derivation was generalised to adiabatic fluids
 in which the entropy on each particle is conserved, and then extended in a
 self-consistent manner to
 incorporate the effects of viscous heating. Other non-adiabatic effects
 such as molecular cooling and radiative diffusion may be easily incorporated
 in a similar way, ensuring that the net rate of change of energy in
 the system due to such processes is equal to the time rate of
 change of the system Hamiltonian.

 The initial set of numerical tests presented here were designed to allow a
 direct comparison between a standard formulation of SPH employing spatially
 and temporally varying smoothing lengths, and our formulation of SPH in
 which the so-called $\nabla h$ terms are included. It is apparent from the
 results of these calculations that the inclusion of the $\nabla h$ terms has
 no detrimental effects on the ability of the algorithm to reproduce,
 with reasonable accuracy, the
 qualitative and quantitative features of known problems.
 In fact, owing to the improved energy conservation, it is found that
 certain features of the adiabatic collapse of a cold gas sphere are more
 accurately modeled when the $\nabla h$ terms are included.

 The advantages of including the $\nabla h$ terms in SPH are, however, most
 starkly brought to light by the numerical calculations specifically designed
 to test the conservation of energy. It is obvious from Figs. (12.) and
 (13.) that including the $\nabla h$ terms leads to a dramatic
 improvement in conservation properties when the entropy equation (\ref{Eent})
 is integrated. Although total energy is formally conserved when the
 internal energy equation (\ref{uinteq1}) is used and the $\nabla h$ terms are
 neglected in equations (\ref{gradp5}) and (\ref{uinteq1}), it is simple to
 show that entropy is no longer conserved when modeling adiabatic flows
 (e.g. Hernquist 1993). If the $\nabla h$ terms are included, however, then
 equations (\ref{Eent}) and (\ref{uinteq1}) can be shown to be strictly
 equivalent, leading to the conservation of entropy and energy in both cases.
 Computational advantage may be obtained by choosing to
 integrate the entropy equation (\ref{Eent}) since the use of velocities
 that are not properly time-centred in (\ref{Eent}) still results in
 excellent energy conservation throughout the bulk of the flow.
 Test calculations
 summarised in Table 3. imply that the problem of non-conservation is
 not due to gravitational, viscous, or discretisation effects, but is
 due to the time variability of the $h_i$s driving the evolution of the
 total energy.

 Up until now, discussions concerning $\nabla h$ terms have
 tended to view
 the role of these additional terms as providing information about the
 local spatial variability of the smoothing lengths with respect to
 one another. However, once a functional
 form for the smoothing lengths has been defined in terms of inter-particle
 distances, as in equation (\ref{hk}), this interpretation appears not to be
 correct. Instead, it is apparent from equation (\ref{gradp5}) that
 the $\nabla h$ terms
 provide information about the way $\nabla W({\bf r}_{ij}, h_i)$ changes
 when we make changes to $h_i$, independently of the way in which changes
 are made to the smoothing lengths of neighbouring particles. Thus,
 it is no longer obvious that one may justify neglecting $\nabla h$ terms
 on the grounds that local spatial variations of the smoothing lengths are
 likely to be small because quantities do not vary greatly over the
 scale of the $h_i$s. Instead, the $\nabla h$ terms are accounting for the
 local time variability of the smoothing lengths, as the particles change their
 relative positions, and as such can only
 be neglected when the time scale over which the smoothing lengths vary
 appreciably is longer than both the dynamical time scale,
 and the time period over which
 the integration is performed.
 This is often not the case in calculations performed with SPH such as
 the collapse of gas clouds, or stellar collisions. It is for the above reasons
 that the calculation performed by Hernquist (1993) in which each particle
 was given its own individual smoothing length, which was {\em not} allowed
 to vary in time, showed no substantial errors in energy conservation.
 $\nabla h$ terms should not arise in such a situation since, in the
 Lagrangian sense, the smoothing
 lengths are not then functions of the particle positions/spatial coordinates.

 Finally, although it is relatively
 simple to derive a set of conservative
 equations from the Hamiltonian in SPH, taking into account the variability of
 the smoothing lengths, the practical problem of defining a functional
 form for the $h_i$s still remains. An important consideration is
 computational expense. If too many particles are allowed to contribute to
 the calculation of the smoothing lengths then the scheme becomes too
 costly in terms of CPU time. In order to avoid this problem, we have
 allowed only one particle to contribute to each $h_i$, as in equation
 (\ref{hi}), for most of the calculations in this paper.
 At first glance this approach seems a little dangerous because
 additional fluctuations are then introduced into the pressure forces.
 In fact, by inspecting equation (\ref{gradp5}) one might expect the $\nabla h$
 terms to be of the same order as the usual force term, in which case the
 fluctuations  would be very large. However, the quantity ${\partial W} /
 {\partial h}$ has a change of sign, with $\partial W/ \partial h < 0$ if
 $0<u/h < 1$, and $\partial W/ \partial h > 0$ for $1<u/h <2$. This leads to
 a process of self-cancellation, resulting in a much more
 diminished contribution
 from the $\nabla h$ terms than one may have initially expected.
 A problem may still exist for particles that do not play the role of
 $imax$ for other particles in the system, or for particles that play
 the role of $imax$ for too many particles. In such a situation,
 the $\nabla h$ contribution to the pressure force becomes
 either uni-directional, or
 else is of large magnitude due to the additive nature of these terms. For a
 purely random distribution of particles, it would be expected that
 such a situation might occur quite regularly. In practice, however, it
 seems that the system undergoes a process of relaxation in which
 correlations are set up between the particles, so that such occurrences
 are reduced to a few rogue incidents that have little effect on the
 overall evolution of the system. This is akin to the process
 described by Monaghan (1988) in order to explain why error
 estimates based on Monte-Carlo theory are larger that those
 observed in practice from SPH calculations.
 The test calculations presented in Section (6) show that in
 general these fluctuations are indeed small, and do not lead to an increase in
 the numerical diffusivity of SPH. In fact, the inclusion of $\nabla h$ terms
 reduces the numerical diffusion in an oscillating polytrope
 due to improved energy conservation. However, we would still advise caution
 when using the functional form (\ref{hi}) for calculating the smoothing
 lengths, and suggest that any effects that may arise from its use be studied
on
 a case by case basis.
 Should the increased pressure force fluctuations become a cause for concern,
 then one can easily increase the number of particles contributing to the
$h_i$s
 through equation (\ref{hn6}), though at the cost of increased CPU time and
 more complicated book-keeping. However, it seems from Fig. (14.c) that
 only a small number
 of particles ($\approx 6$) are required to all but remove the excess
 fluctuations.

 \vspace{5mm}

 \noindent {\bf Acknowledgements.}
  We thank David Balding, Mark Beaumont, and Diana Pallant
  of the QMWC Statistics group for generously allowing us extensive use of
  their computer facilities.
  This work was supported by the
  SERC QMW rolling theory grant (GR/H/09454).
  RN is supported by an SERC studentship.
  \pagebreak

 \newpage

 \large
 \begin{center}
 \noindent {\bf References.}
 \end{center}
 \normalsize

 \begin{trivlist}
 \item[] Barnes, J., Hut, P., 1986, \nat 324, 446
 \item[] Cox, J.P., 1980, Theory of Stellar Pulsation, Princeton Univ. Press.
 \item[] Evrard, A.E., 1988, \mnr  235, 911
 \item[] Gingold, R.A., Monaghan J.J., 1977, \mnr  181, 375
 \item[] Gingold, R.A., Monaghan J.J., 1982, {\it J. Comp. Phys.}  46,
 429
 \item[] Goodman, J., Hernquist, L., 1991, \apj 378, 637
 \item[] Hernquist, L., 1987, \apjs  64, 715
 \item[] Hernquist, L., 1993, \apj  404, 717
 \item[] Hernquist, L., Katz, N., 1989, \apjs  70, 419
 \item[] Lucy, L.B., 1977, \aj  83, 1013
 \item[] Monaghan, J.J., 1988, {\it Comp. Phys. Comm.} 48, 89
 \item[] Monaghan, J.J., 1992, \anrev 30, 543
 \item[] Monaghan, J.J., Gingold, R.A., 1983, {\it J. Comp. Phys.}  52, 374
 \item[] Monaghan, J.J., Lattanzio, J.C., 1985 \ana  149, 135
 \item[] Nelson, R.P., Papaloizou, J.C.B., 1993, \mnr 265, 905
 \item[] Sod, G.A., 1978, {\it J. Comp. Phys.}, 27, 1
 \item[] Thomas, P., 1987, PhD thesis, Cambridge University.
 \end{trivlist}

 \newpage

 \begin{table}[h]
 \caption{Colliding Polytropes.}
 \begin{center}
 \begin{tabular}{llllllr}  \hline \hline
  Run & $\nabla h$ terms & Equation & ${\cal N}_{TOL} $
 & $E$- deviation ($\%$)& \\ \hline
     &       &    & &           &               \\
$\; C1$ & Included & Entropy  & 50 & $\;\;\;\;\;\;\ 0.8 \%$ & \\
        &       &    & &           &               \\
 $\; C2$ & Included &  Entropy  & $ 50 \pm 1$   & $\;\;\;\;\;\;\  2.3\%$  &  \\
        &       &    &  &         &               \\
 $\; C3$ & Included &  Entropy & $ 50 \pm 2 $ & $\;\;\;\;\;\;\ 3.2\%$  & \\
        &       &    &   &        &               \\
 $\; C4$ & Excluded & Entropy &$ 50 \pm 2$  & $\;\;\;\;\;\;\ 9.8 \%$    & \\
        &       &    &   &        &               \\
 $\; C5$ & Included & Energy &$ 50 \pm 2 $ & $\;\;\;\;\;\;\  1.8\%$     & \\
        &       &    &   &        &          \\
 $\; C6$ & Excluded & Energy &$ 50 \pm 2 $ & $\;\;\;\;\;\;\  2.1\%$     & \\
        &       &    &   &        &          \\
 \hline \hline
 \end{tabular}
 \end{center}
 \end{table}

 \newpage

 \begin{table}[h]
 \caption{Expanding Polytropes.}
 \begin{center}
 \begin{tabular}{llllllr}  \hline \hline
  Run & $\nabla h$ terms & Equation & ${\cal N}_{TOL} $
 & $ E$- deviation ($\%$)& \\ \hline
&       &    & &           &               \\
$\; E1$ & Included & Entropy  & 50 & $\;\;\;\;\;\;\ 0.01 \%$ & \\
        &       &    & &           &               \\
 $\; E2$ & Included &  Entropy & $ 50 \pm 2 $ & $\;\;\;\;\;\;\ 0.15\%$  & \\
        &       &    &   &        &               \\
 $\; E3$ & Excluded & Entropy &$ 50 \pm 2$  & $\;\;\;\;\;\;\ 1.48 \%$    & \\
        &       &    &   &        &               \\
 $\; E4$ & Excluded & Entropy &$ 50 $  & $\;\;\;\;\;\;\ 1.48 \%$    & \\
        &       &    &   &        &               \\
$\; E5$ & Included & Energy &$ 50 \pm 2 $ & $\;\;\;\;\;\;\  0.13\%$     & \\
        &       &    &   &        &          \\
 $\; E6$ & Excluded & Energy &$ 50 \pm 2 $ & $\;\;\;\;\;\;\  0.13\%$     & \\
        &       &    &   &        &          \\
 \hline \hline
 \end{tabular}
 \end{center}
 \end{table}

 \newpage

 \begin{table}[h]
 \caption{Timing Tests for Colliding Polytropes: N=4170}
 \begin{center}
 \begin{tabular}{lllllr}  \hline \hline
  $\nabla h$ terms & $h_i$-eqn & ${\cal N}_{TOL} $
 & Time/t-step(s)& \\ \hline
&       &     &           &               \\
  Included & $\;\;$ \ref{hi}  & 50 & $\;\;\;\;\ \ \ \ 95  $ & \\
        &           & &           &               \\
  Included & $\;\;$  \ref{hi} & $ 50 \pm 1 $ & $\;\;\;\;\ \ \ \ 81 $   & \\
        &           &   &        &               \\
  Included & $\;\;$ \ref{hi} &$ 50 \pm 2$  & $\;\;\;\;\ \ \ \ 78 $   & \\
        &           &   &        &               \\
  Excluded & $\;\;$ \ref{hi} &$ 50 \pm 2 $ & $\;\;\;\;\ \ \ \  72 $   & \\
        &           &   &        &          \\
  Included & $\;\;$ \ref{hn6} &$ 50 \pm 1 $ & $\;\;\;\;\ \ \ \  95 $   & \\
        &           &   &        &          \\
  Included & $\;\;$ \ref{hn6} &$ 50 \pm 2 $ & $\;\;\;\;\ \ \ \  87 $   & \\
        &           &   &        &          \\
 \hline \hline
 \end{tabular}
 \end{center}
 \end{table}

 \newpage

 \begin{center}
 {\bf Table  Captions.}
 \end{center}

 \noindent {\bf Table 1.} Results from colliding polytropes. Column(2)
 identifies whether the $\nabla h$ terms were included, column(3)
 whether the entropy or internal energy equation was integrated.
 Column(4) gives the number of nearest neighbours within 2$h_i$, and
 column(5) the percentage deviation from perfect energy conservation.

 \vspace{3mm}

 \noindent {\bf Table 2.} Results of timing tests performed for calculations of
 colliding polytropes. Column (1) indicates whether $\nabla h$ terms were
 included. Column (2) describes which equation, (\ref{hi}) or (\ref{hn6}),
 was used to calculate smoothing lengths. The values taken for ${\cal N}_{TOL}$
 are given in column (3), and the CPU per time step, measured in seconds, is
 given in column (4).

 \vspace{3mm}

 \noindent {\bf Table 3.} Results from expanding polytropes. Column(2)
 identifies whether the $\nabla h$ terms were included, column(3)
 whether the entropy or internal energy equation was integrated.
 Column(4) gives the number of nearest neighbours within
 2$h_i$, ${\cal N}_{TOL}$, and
 column(5) the percentage deviation from perfect energy conservation.

 \newpage

 \begin{center}
 {\bf Figure Captions.}
 \end{center}

 \vspace{3mm}

 \noindent{\bf Figure 1.} (a) Density, (b) pressure, (c) velocity, (d)
 entropic profiles in a one-dimensional shock tube, using
 artificial viscosity defined by equation (\ref{Pi}). Analytical solution
 is shown by dashed lines. In this case
 the $\nabla h$ terms were neglected.

 \vspace{3mm}

 \noindent{\bf Figure 2.} (a) Density, (b) pressure, (c) velocity, (d)
 entropic profiles in a one-dimensional shock tube, using
 artificial viscosity defined by equation (\ref{Pi}). In this case
 the $\nabla h$ terms were included.

 \vspace{3mm}

 \noindent{\bf Figure 3.} Density profile during adiabatic collapse of
 a gas sphere, initially having an isothermal energy distribution and
 a $1/r$ density profile. Following Evrard (1988), density is measured
 in units of $\rho_* = 3M_T/4\pi R^2$, where $R$ is initial radius
 and $M_T$ is total mass. Dimensionless time, normalised to free-fall
 time at outer radius, is shown in top right corner
 of each frame. In this case $\nabla h$ terms were neglected.

 \vspace{3mm}

 \noindent{\bf Figure 4.} Thermal energy distribution, normalised to
 $u_* = GM_T/R$, during adiabatic collapse of gaseous sphere. Dimensionless
 time is shown in top right corner of each frame. $\nabla h$ terms were
 neglected.

 \vspace{3mm}

 \noindent{\bf Figure 5.} Pressure during adiabatic collapse of gas sphere,
 normalised to $P_* = \rho_* u_*$. Dimensionless time is shown at
 top right corner of each frame. $\nabla h$ terms were neglected.

 \vspace{3mm}

 \noindent{\bf Figure 6.} Radial velocity profiles during collapse of gas
sphere
 described in text, normalised to $v_* = (GM_T/R)^{(1/2)}$. Dimensionless
 time is shown in top left corner of each frame. $\nabla h$ terms were
 neglected.

 \vspace{3mm}

 \noindent{\bf Figure 7.} Same as Fig. 3, except that $\nabla h$ terms were
 included.

 \vspace{3mm}

 \noindent{\bf Figure 8.} Same as Fig. 4, except that $\nabla h$ terms were
 included.

 \vspace{3mm}

 \noindent{\bf Figure 9.} Same as Fig. 5, except that $\nabla h$ terms were
 included.

 \vspace{3mm}

 \noindent{\bf Figure 10.} Same as Fig. 6, except that $\nabla h$ terms were
 included.

 \vspace{3mm}

 \noindent{\bf Figure 11.} Head on collision between identical
 $n=3/2$ polytropes,
 each containing 2085 particles. Time is indicated at the top right corner
 of each panel, measured in computational units. Initial velocities
 and set-up is described in the text.

 \vspace{3mm}

 \noindent{\bf Figure 12.} Evolution of total energy in simulations of
 head on collisions between identical $n=3/2$ polytropes.
 The calculations represented in above figure correspond to the following
 runs listed in Table 1:
 $C1$ -- Solid line; $C2$ -- Dashed line; $C3$ -- Dot-dash-dotted line; $C4$ --
 Dotted line; $C6$ -- Dash-dot-dot-dot-dashed line.  Units are in computational
 units.

 \vspace{3mm}

 \noindent{\bf Figure 13.}  Evolution of total energy in simulations of
 freely expanding polytropes in absence of gravity and viscosity. Units
 are in computational units. The calculations represented correspond to
 the following runs listed in Table 2:
 $E1$ -- Solid line; $E2$ -- Dashed line; $E3$ -- Dash-dot-dashed line;
 $E6$ -- Dotted line.

 \vspace{3mm}

 \noindent{\bf Figure 14.} The graphs represent the pressure force,
 experienced as
 a function of time, by a single particle during free expansion
 of an $n=3/2$ polytrope. (a) $\nabla h$ terms are not included in
 equations of motion. (b) $\nabla h$ terms are included and smoothing
 lengths are calculated according to equation (\ref{hi}). (c) $\nabla h$
 terms are included and smoothing lengths are calculated according to
 equation (\ref{hn6}), with ${\cal N}_{far}=6$. Units are in computational
 units.

 \vspace{3mm}

 \noindent{\bf Figure 15.} The graphs represent the various energies,
 as functions
 of time, during the normal mode oscillations of an $n=3/2$ polytrope.
 In each panel, the line styles representing each of the various energies are:
 dashed - thermal energy; dotted - kinetic energy; solid - total energy
 (including energy dissipated by viscosity); dash-dot-dashed - gravitational
 potential energy.
 Units are in computational units.
 (a). $\nabla h$ terms included. (b). $\nabla h$ terms neglected.

\end{document}